\documentclass[twoside,12pt]{article}
\usepackage{epsfig}

\def\ltsima{$\; \buildrel < \over \sim \;$}
\def\simlt{\lower.5ex\hbox{\ltsima}}
\def\gtsima{$\; \buildrel > \over \sim \;$}
\def\simgt{\lower.5ex\hbox{\gtsima}}

\newcommand{\be}{\begin{equation}}
\newcommand{\ee}{\end{equation}}
\newcommand{\bea}{\begin{eqnarray}}
\newcommand{\eea}{\end{eqnarray}}

\begin{document}

\title{
The BOOMERanG experiment \\ and the curvature of the Universe }

\author{S. Masi$^{1}$, P.A.R.~Ade$^{2}$,
J.J.~Bock$^{3}$, J.R.~Bond$^{4}$, J.~Borrill$^{5}$, \\
A.~Boscaleri$^{6}$, K.~Coble$^{7}$,
C.R.~Contaldi$^{4}$, B.P.~Crill$^{8}$, P.~de Bernardis$^{1}$ \\
G.~De~Gasperis$^{9}$,  G.~De~Troia$^{1}$, P.~Farese$^{7}$,
K.~Ganga$^{10}$, \\
M.~Giacometti$^{1}$, E.~Hivon$^{10}$, V.V.~Hristov$^{8}$,
A.~Iacoangeli$^{1}$, A.H.~Jaffe$^{11}$, \\
W.C.~Jones$^{8}$, A.E.~Lange$^{8}$, L.~Martinis$^{12}$,
P.~Mason$^{8}$, P.D.~Mauskopf$^{2}$, \\
A.~Melchiorri$^{13}$, P.~Natoli$^{9}$, T.~Montroy$^{7}$,
C.B.~Netterfield$^{14}$, \\
E.~Pascale$^{6}$, F.~Piacentini$^{1}$, D.~Pogosyan$^{4}$,
G.~Polenta$^{1}$, F.~Pongetti$^{15}$, \\
S.~Prunet$^{4}$, G.~Romeo$^{15}$,
J.E.~Ruhl$^{7}$, F.~Scaramuzzi$^{12}$, N.~Vittorio$^{10}$ \\
\\
$^{1}$ Dipartimento di Fisica, Universit\'a La Sapienza, Roma. \\
$^{2}$ Dept. of Physics and Astronomy, Cardiff.
$^{3}$ Jet Propulsion Laboratory, Pasadena. \\
$^{4}$ C.I.T.A., University of Toronto. $^{5}$ N.E.R.S.C., LBNL, Berkeley. \\
$^{6}$ IROE-CNR, Firenze, Italy. $^{7}$ Dept. of Physics, Univ. of California, \\
Santa Barbara. $^{8}$ California Institute of Technology, Pasadena. \\
$^{9}$ Department of Physics, Second University of Rome. \\
$^{10}$ IPAC, Caltech, Pasadena. $^{11}$ CFPA, Berkeley. \\
$^{12}$ ENEA, Frascati, Italy. $^{13}$ Nuclear and Astrophysics
Laboratory, Oxford. \\
$^{14}$ Depts. of Physics and Astronomy, University of Toronto. \\
$^{15}$ Istituto Nazionale di Geofisica, Roma. }

\maketitle

\begin{abstract} We describe the BOOMERanG experiment and its main
result, i.e. the measurement of the large scale curvature of the
Universe. BOOMERanG is a balloon-borne microwave telescope with
sensitive cryogenic detectors. BOOMERanG has measured the angular
distribution of the Cosmic Microwave Background on $\sim 3\%$ of
the sky, with a resolution of $\sim 10$ arcmin and a sensitivity
of $\sim 20 \mu K$ per pixel. The resulting image is dominated by
hot and cold spots with rms fluctuations $\sim 80 \mu K$ and
typical size of $\sim 1^o$. The detailed angular power spectrum of
the image features three peaks and two dips at $\ell =
(213^{+10}_{-13}), (541^{+20}_{-32}), (845^{+12}_{-25}  )$ and
$\ell = (416^{+22}_{-12}), (750^{+20}_{-750})$, respectively. Such
very characteristic spectrum can be explained assuming that the
detected structures are the result of acoustic oscillations in the
primeval plasma. In this framework, the measured pattern
constrains the density parameter $\Omega$ to be $0.85 < \Omega <
1.1$ (95\% confidence interval). Other cosmological parameters,
like the spectral index of initial density fluctuations, the
density parameter for baryons, dark matter and dark energy, are
detected or constrained by the BOOMERanG measurements and by other
recent CMB anisotropy experiments. When combined with other
cosmological observations, these results depict a new, consistent,
cosmological scenario.
\end{abstract}
\section{Introduction}

The almost isotropic Cosmic Microwave Background (CMB) accounts
for most of the photons present in our Universe. These photons
have been produced in the very early  Universe, and were last
scattered by free electrons at recombination, about 300000 years
after the Big Bang. After that, CMB photons travel basically
undisturbed along spacetime geodesics for $\sim$ 15 Gyr, reaching
our telescopes redshifted by a factor $\sim 1000$ due to the
expansion of the universe.

Thus, when we look to the image of the CMB, we see the result of
early processes (e.g. the generation of density perturbations at
$t \sim 10^{-30} s$ after the Big Bang, and matter-antimatter
annihilations, at $t \sim 10^{-5}$ s), of processes in the plasma
era till recombination (e.g. acoustic oscillations of the
matter-photons plasma till $t \sim 300000 yrs$), and of the large
scale geometry of the universe (affecting the geodesics followed
by the photons after recombination, from $t \sim 300000 yrs$ to $t
\sim 15 Gyrs$ ). Any small curvature of the Universe and of the
geodesics would affect significantly the image of the CMB,
magnifying or demagnifying it with respect to the Euclidean case.
For this reason experiments mapping the CMB are very sensitive to
the curvature of the Universe, which, according to General
Relativity, is determined by the average mass-energy density of
the Universe $\rho$.

A parametric, general relativistic theory of the anisotropy of the
CMB has been fully developed in the last 35 years, based on the
main cosmological observations: the isotropic expansion of the
Universe, the primordial abundances of light elements, the
existence of the CMB and its black-body spectrum, the large scale
distribution of Galaxies. Detailed models and codes are available
to compute the angular power spectrum of the CMB image given a
cosmological model for the generation of density fluctuations in
the Universe, and a set of parameters describing the background
cosmology \cite{Hu}.

The default model is nowadays the inflationary adiabatic one,
where gaussian, adiabatic density fluctuations are generated from
quantum fluctuations of a scalar field present in the very early
Universe, boosted to cosmological scales by the inflation phase at
$E \sim 10^{15} GeV$ \cite{KT1990}. This modification of the
standard theory is needed in order to solve several paradoxes
intrinsic to the standard Hot Big Bang theory. As we will see
below, this theory offers a natural explanation to several
cosmological observations, but the ingredients required to fit the
data are non trivial, requiring the presence of unobserved "dark
matter" and unknown "dark energy".
\begin{figure}[tb]
\begin{center}
\begin{minipage}[t]{16 cm}
\epsfig{file=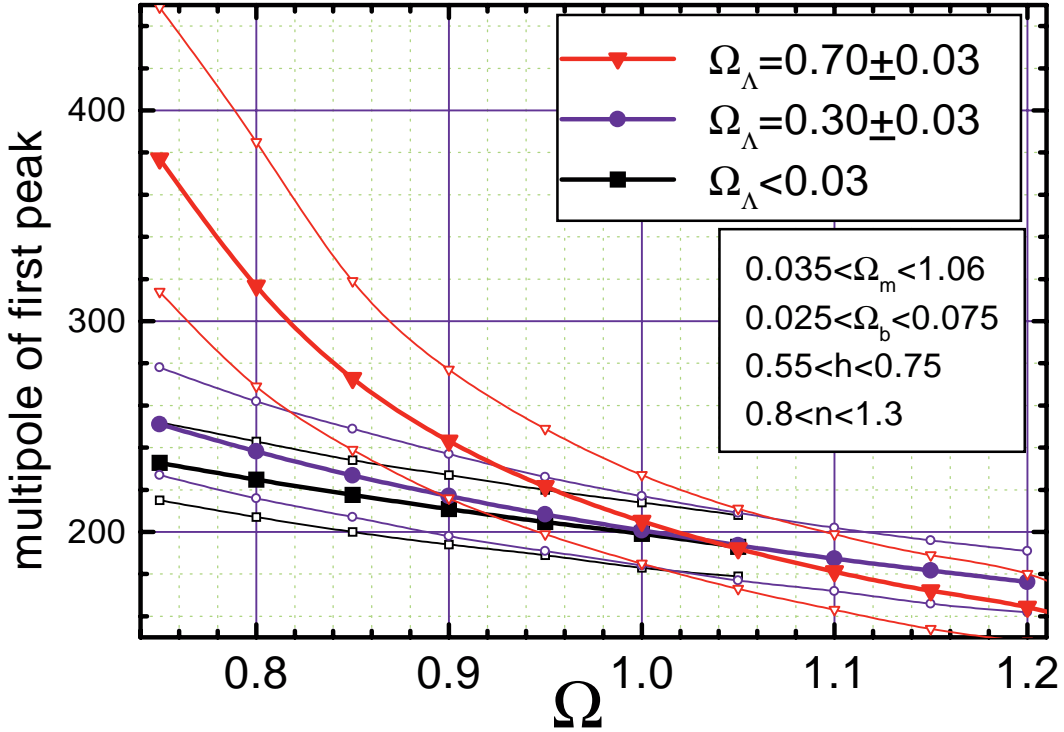,scale=1.0}
\end{minipage}
\begin{minipage}[t]{16.5 cm}
\caption{Position of the first peak in the angular power spectrum
of the CMB as a function of the total matter-energy density
parameter $\Omega$. The results are plotted for a wide database of
adiabatic inflationary models ($\sim 5\times 10^5$ models
corresponding to the ranges of the parameters listed in the bottom
panel), for three sample values of $\Omega_\Lambda$ (as in the top
panel). The thick lines are the average values of $\ell_1$. The
band within thin lines includes all the possible combinations of
the other parameters. See text for details and use of the graph.
\label{fig1}}
\end{minipage}
\end{center}
\end{figure}

The parameters of the model are: the Hubble constant $H_o$
(setting the expansion rate of the Universe); the density
parameter $\Omega$ (i.e. the ratio between $\rho$ and the critical
density $\rho_c = 3H_o^2 / 8 \pi G $); the baryons, dark matter
and dark energy density parameters $\Omega_b$, $\Omega_m$,
$\Omega_\Lambda$; the spectrum of the primordial density
fluctuations generating CMB anisotropy. The latter is usually
expressed as a power law with spectral index $n$.

The general prediction of the model above is a series of "acoustic
peaks" in the angular power spectrum of the CMB. These peaks
derive from acoustic oscillations of the photons-matter plasma
(the primeval fireball).

Density perturbations $\Delta \rho / \rho$ were oscillating in the
primeval fireball as a result of the opposite effects of the
pressure of photons and of gravity. After recombination, photons
pressure becomes unimportant, and $\Delta \rho / \rho$ can grow
and create, through gravitational instability, the hierarchy of
structures we see today in the nearby Universe. There are three
physical processes converting the density perturbations present at
recombination into {\it observable } CMB temperature fluctuations
$\Delta T / T$. They are: the photon density fluctuations
$\delta_\gamma$, which can be related to the matter density
fluctuations $\Delta \rho$ once a specific class of perturbations
is specified; the gravitational redshift of photons scattered in
an over-density or an under-density  with gravitational potential
difference $\phi_r$; the Doppler effect produced by the proper
motion with velocity $v$ of the electrons scattering the CMB
photons. In formulas:
\begin{equation}
\frac{\Delta T}{T}(\vec{n}) \approx \frac{1}{4}\delta_{\gamma r}+
\frac{1}{3} {\phi_r \over c^2} - \vec{n} {\vec{v_r} \over c}
\end{equation}
where $\vec{n}$ is the line of sight vector and the subscript $r$
labels quantities at recombination.

There is an acoustic horizon at recombination, with a linear size
of $\sim 300000$ light-years. At any time $t$ in an infinite
Universe regions separated more than $c t$ are not in causal
contact. This means that causal horizons, with size $ct$, exist in
the Universe. The acoustic horizon has a size similar to the
causal horizon, since in the primeval plasma the speed of sound is
close to the speed of light. Since the horizon expands with time,
any given proper size will eventually become smaller than the
acoustic horizon. The time when the proper size of a perturbation
becomes equal to the acoustic horizon is called horizon crossing.
Perturbations larger than the acoustic horizon at recombination
have never been in acoustic contact before, and have not been able
to oscillate. Perturbations smaller that that size have oscillated
after crossing the horizon, and arrive to recombination with a
phase depending on their intrinsic size. The image of the CMB,
which is directly observable by means of CMB anisotropy
experiments, is a processed image of density perturbations present
at recombination. The size of the acoustic horizon is expected to
be evident in the size distribution of the detected CMB structures
$\Delta T(\alpha, \delta)$. In fact, perturbations with a size
close to the acoustic horizon had just enough time to fully
compress or rarefy before recombination, and will be evident as
cold or hot spots. If we compute the angular power spectrum
$c_\ell$ of the image, where $c_\ell = \langle |a_{\ell, m}|^2
\rangle$, and $\Delta T(\alpha, \delta) = \sum_{\ell, m} a_{\ell,
m} Y^\ell_m (\alpha, \delta)$, we expect to see a peak
corresponding to the angular distance subtended by the acoustic
horizon at recombination. In an Euclidean universe, the subtended
angle is simply 300000 lyr divided by 15 Glyr and multiplied by a
factor 1000, to take into account the subsequent expansion of the
Universe. We thus expect structures with a typical angular scale
$\theta_1 \sim 1^o$, which corresponds to a multipole $\ell_1 \sim
\pi / \theta_1 \sim 200$.

This is correct only if the geometry of the Universe is Euclidean,
not curved, i.e. if the average mass-energy density of the
Universe is the critical one ($\Omega=1$). If, instead, the
mass-energy density is higher than critical ($\Omega
> 1$), the geometry of space will have a positive curvature and
the photons will travel along curved geodesics. The excess density
will act as a magnifying glass, and the same fluctuations in the
CMB will appear as spots larger than $1^o$.  The opposite will
happen if the density is lower than critical, acting as a
de-magnifying glass and producing a typical angular size of the
fluctuations smaller than $1^o$. By measuring the location of the
peak it will thus be possible to measure $\Omega$. The
quantitative treatment of this angular-size vs distance test can
be found in \cite{Wein2000}, \cite{Mel00}. In general, $\ell_1$
decreases when $\Omega$ increases, but the location of the first
peak is also controlled by $\Omega_\Lambda$, which effectively
changes our distance from recombination (see fig.1). Only if
$\Omega_\Lambda =0$ the simple relationship $\ell_1 \sim \Omega^{-
{1\over 2}}$ holds \cite{Wein2000}.

CMB experiments are now starting to measure the angular power
spectrum with  sufficient accuracy to infer $\Omega$ and several
other cosmological parameters. This inverse procedure, i.e.
measure the cosmological parameters given the observed angular
power spectrum and its measurement error, is a complex one, and is
further complicated by the presence of degeneracies between the
cosmological parameters. Different combinations of the parameters
can generate the same power spectrum of the CMB \cite{Efs99}, so
priors (coming from independent cosmological evidence) must be
assumed in the analysis \cite{Lan01}. The measurement of curvature
is quite robust in this sense. In fig.1 it is evident that a
measurement of e.g. $\ell_1 = 200$ is consistent only with models
with $0.85<\Omega<1.1$, for any possible value of the other
parameters $\Omega_\Lambda, \Omega_b, \Omega_m, h, n$. More
precise measurements of $\Omega$ can be obtained from a full
likelihood analysis of the power spectrum data (see below). Once
the default model and rather weak priors (for example on the
Hubble constant) are assumed, the measurements set very strong
constraints also on all the other parameters, and the results are
consistent with independent cosmological observations (see e.g.
\cite{Teg01}. A new "precision phase" in the cosmological research
seems to be starting. This, however, has a cost, which is the
introduction of "strange" components in the Universe like dark
matter and dark energy. Only further observations will show if the
latter are just artifices, like Ptolemy's epicycles deferents and
late additions, or, instead, are very important new discoveries.

In this paper we show how the curvature of the Universe has been
measured by BOOMERanG. We describe the experiment, then the
experimental strategy and observations obtained in the long
duration flight, and finally the cosmological implications.

\section{The BOOMERanG experiment}

\subsection{\it The Observable is small }

The CMB is a faint, almost isotropic glow of microwaves. Its
purely Planck spectrum has been measured with great accuracy by
the COBE-FIRAS experiment \cite{Math90}, \cite{Math99}. The
temperature of the CMB is $T_o = 2.725 K$. Anisotropy in the CMB
has been first detected at large angular scales ($>7^o$) by the
COBE-DMR experiment \cite{Smoot1992}. At large angular scales the
level of the fluctuations detected by COBE-DMR is $\sim$ 10 parts
per million of the average level of the CMB: $\Delta T _ {rms}
\sim 30 \mu K$ \cite{Ban96}. This measurement does not allow to
measure the curvature of the Universe, since the angular
resolution is not enough to resolve the degree-sized hot and cold
spots. But it allows us to normalize the angular power spectra
computed theoretically, and to predict that the rms of the
fluctuations detected with a resolution of, say, 0.2$^o$, will be
in the $100 \mu K$ range. This means that a very sensitive
microwave telescope is needed. The detection of a sub-degree
resolution map of the CMB represents a formidable experimental
challenge, since the emissions of the telescope, of the earth
atmosphere, of astrophysical foregrounds, and of the isotropic
component of the CMB itself, are much higher.

Atmospheric emission is reduced by flying the experiment above the
bulk of earth's atmosphere (38 km), by means of a stratospheric
balloon. The observation frequencies of the BOOMERanG experiment
(90, 150, 240, 410 GHz) have been carefully selected in order to
work in the range where residual atmosphere and galactic
foregrounds are minimum. In fig.2 we compare the spectrum of CMB
anisotropy to the spectrum of different foregrounds and to the
observation frequencies of the BOOMERanG multi-band photometer.
\begin{figure}[tb]
\begin{center}
\begin{minipage}[t]{16 cm}
\epsfig{file=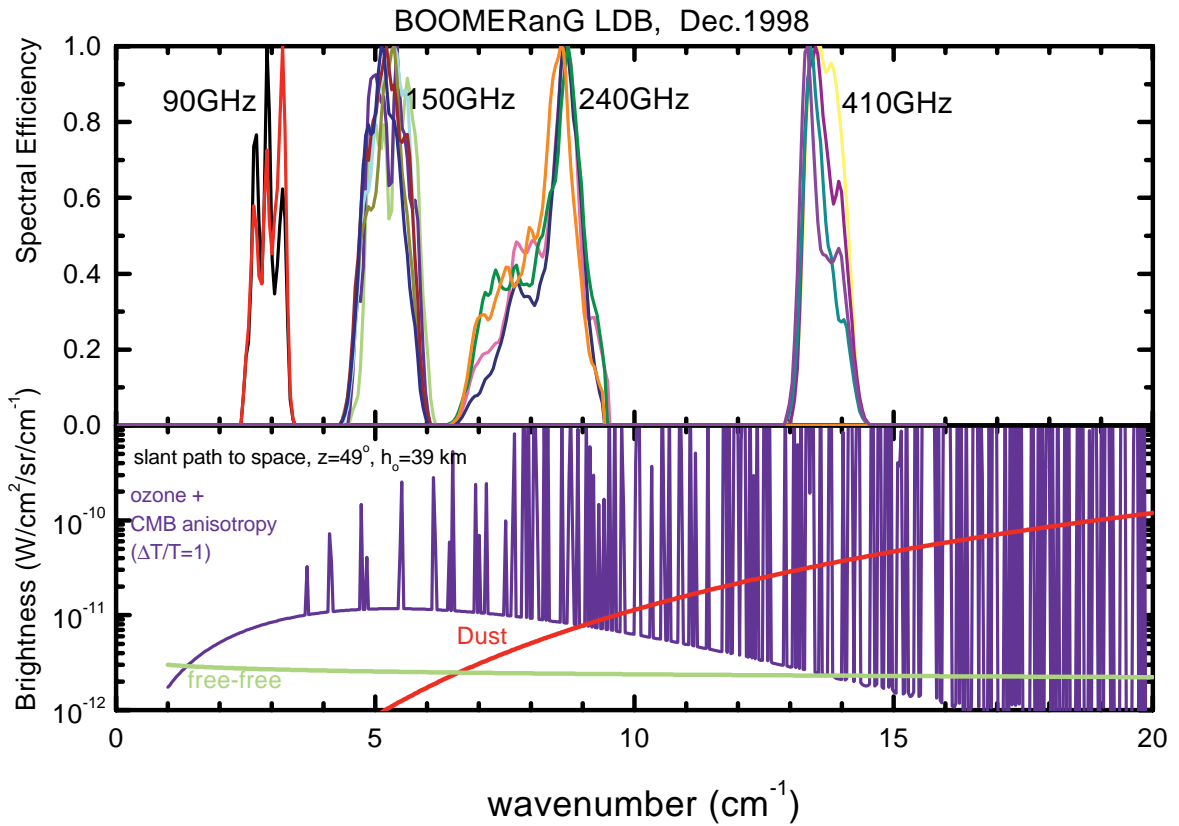,scale=1.0}
\end{minipage}
\begin{minipage}[t]{16.5 cm}
\caption{Upper panel: spectral efficiency of all the 16 detection
channels of the BOOMERanG experiment. Lower panel: spectrum of the
CMB anisotropy, compared to the typical spectra of ozone emission
at stratospheric altitudes, of diffuse interstellar dust, of
interstellar free-free. \label{fig2}}
\end{minipage}
\end{center}
\end{figure}
CMB anisotropy brightness has been computed as
\begin{equation}
\Delta I (\nu) = {x e^x \over e^x - 1 } B(\nu , T_o) {\Delta T
\over T_o}
\end{equation}
with $x = h \nu / k T_o$, where $B(\nu , T_o)$ is the Planck
function, and assuming $\Delta T / T_o = 1$ in place of $\sim
10^{-4}$. The atmospheric emission at balloon altitude has been
computed for a slant path to space with elevation of 41$^o$, using
the HITRAN atmospheric model. Residual water vapor is very low,
and oxygen emission is very isotropic. We have included only
ozone, which is known to be distributed in stratospheric clouds,
and can be potentially dangerous for these measurements
\cite{Pag90}. The two bands sensitive to CMB anisotropy are the 90
and 150 GHz ones, which are close to the peak of the CMB
anisotropy brightness. The 240 and 410 GHz bands are used to
monitor atmospheric and interstellar emission in order to control
the level of contamination in the lower frequency maps.

\subsection{\it Experimental approach }

The telescope scans the sky in order to separate the anisotropic
CMB signal from any isotropic foreground (like instrumental
emission) and background (like the isotropic component of the
CMB). The signals from the detectors are AC-coupled, so that
isotropic emission is rejected. The sky scan is performed at
constant elevation. This simple solution strongly reduces the
atmospheric signal (only gradients and inhomogeneities contribute,
see below).

The azimuth scan speed ${\dot A}$ is between 1 and 2 deg/s.
Different multipoles $\ell$ in the CMB are converted into
different sub-audio frequencies $f$ in the detectors: $f = \ell
{\dot A} / \pi \cos e $ where $e$ is the elevation of the beam.
Multipoles between 20 and 1000 are converted into frequencies $f$
between 0.15-0.3 and 7.5-15 Hz. This range lies inside the flat
response band of the detectors, above their 1/f knee frequency,
and above the characteristic pendulation frequency of the balloon
and payload system.

The payload is flown around Antarctica aboard of a long duration
balloon (LDB, $>7$ days) performed by NASA-NSBF. The payload
drifts by thousands of km, circumnavigating the Antarctic
continent at a nearly constant latitude of -79$^o$. This enables
long integrations, wide sky coverage and extensive tests for
systematic effects, through the repetition of the measurements
under different experimental conditions. Since the measurements
are obtained from different locations we can control spillover
from ground, ice and sea. The consistency between maps obtained
from different locations is the best test against the presence of
sidelobes contributions in the measurements.

The long duration of the flight allows to control the contribution
of the sun in the far sidelobes.

"day" (sun at high elevations) vs "night" (sun at low elevations)
observations have different scan directions on the same area,
producing nicely cross-linked scans, useful to reconstruct the map
of the sky.

We select the lowest foreground sky region for the observations.
This happens to be far from the sun in the antarctic summer
(constellations of Caelum, Doradus, Pictor, Columba, Puppis).
Performing 60$^o$ wide azimuth scans, and tracking the azimuth of
the center of the best sky region with the center of the scan, we
can cover the best $\sim 4 \%$ of the sky in one full day of
observation, and repeat the measurement for many days.

\subsection{\it Detection technique}

At these frequencies the most sensitive detectors for continuum
radiation are cryogenic bolometers. A bolometer consists of a
broadband absorber with heat capacity, C, that has a weak thermal
link, G, to a thermal bath at a temperature, T$_{\rm b}$. Incident
radiation produces a temperature rise in the absorber that is read
out with a current biased thermistor. The sensitivity of a
bolometer expressed in Noise Equivalent Power (NEP) is given by:
\begin{equation}
{\rm NEP_{bolo}} = \gamma \sqrt{4k{\rm T_b}^2 {\rm G}}
\end{equation}
where $\gamma$ is a constant of order unity that depends weakly on
the sensitivity of the thermistor.  The need for a cryogenic
temperature of the bolometer is evident. Bolometers have been
operated at temperatures as low as 0.05 K. The dynamical equation
for the temperature of the thermistor can be expressed as:
\begin{equation}
\frac{\rm dT_{bolo}}{\rm dt} = \frac{\rm
P_{in}-G(T_{bolo}-T_b)}{\rm C}
\end{equation}
From this equation it is evident that incident background power is
equivalent to physical temperature of the device: both need to be
as low as possible to get the best NEP. In this respect,
balloon-borne experiments allow to achieve a very low background
power on the detectors, with the contributions from the CMB, the
residual atmosphere and the thermal emission of the optical system
are similar. From the same equation we can see that bolometers
have a finite bandwidth, limited by the time for the absorber to
come to equilibrium after a change in incident power, $\tau =$
C/G. Previous balloon-borne bolometric receivers have been limited
in sensitivity or bandwidth by the properties of the materials
used for fabrication of the detectors.

Bolometers are also limited in sensitivity by external sources of
noise such as cosmic rays, microphonic disturbances, and radio
frequency interference (RFI). Of particular importance for
BOOMERanG is the cosmic ray rate from balloon altitude in the
Antarctic, which is about an order of magnitude higher than from
temperate latitudes, because the magnetic field of the earth
funnels charged particles to the poles.

The new technology of composite bolometers with metallized
Si$_3$N$_4$ spider-web absorbers and Ge thermistors \cite{Maus97,
Bock99} has produced bolometers with NEP as low as $2 \times
10^{-17} W/\sqrt{Hz}$ at a physical temperature of the device of
300 mK.  These bolometers have lower heat capacity, lower thermal
conductivities, lower cosmic ray cross section, and less
sensitivity to microphonic heating than previous 300 mK
bolometers. They also feature very low 1/f noise, with a knee
frequency below 0.01 Hz. Spider web bolometers optimized for 100
mK operation and have been recently selected for the HFI
instrument on ESA's CMB mission Planck, which uses a scan strategy
similar to BOOMERanG.

The BOOMERanG detectors are optimized to have the maximum
sensitivity possible under the estimated loading conditions during
the stratospheric flight. The achieved in-flight sensitivities are
reported in table 1.

Most bolometric detectors in use today employ a high-impedance
semiconductor thermistor biased with a constant current. JFET
preamplifiers have been used to provide a combination of low
voltage and current noise well matched to typical bolometer
impedances, but exhibit excess voltage noise at frequencies
typically below a few Hz and have limited the achievable bandwidth
of these DC biased bolometers.  At lower frequencies, drifts in
the bias current, drifts in the temperature of the heat sink, and
amplifier gain fluctuations have been expected to limit the
ultimate stability of single bolometer systems.

\begin{figure}[tb]
\begin{center}
\begin{minipage}[t]{16 cm}
\epsfig{file=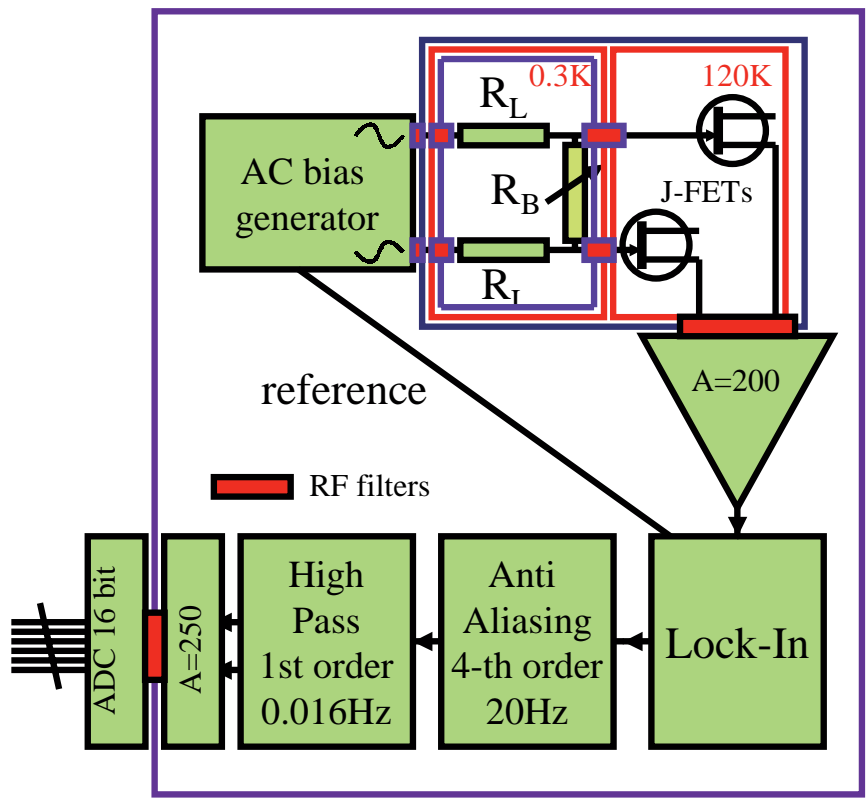,scale=1}
\end{minipage}
\begin{minipage}[t]{16.5 cm}
\caption{Block diagram of the readout electronics for BOOMERanG.
The system is fully differential. Each bolometer is biased through
symmetrical smoothed square-waves at a frequency $\sim 300 Hz$,
higher than its thermal high frequency cut-off. The signal is
readout through a differential JFET follower at 120K, located
inside the cryostat close to the bolometers. The signal is then
amplified and demodulated through a synchronous demodulator locked
to the bias frequency. Anti-alias low pass (20 Hz) and background
removal high-pass (16 mHz) filtering are then applied. The total
amplification is $5\times 10^4$. The amplified signal exit the
RF-enclosure through RFI filters before A-D conversion (16 bit,
62.5 Hz). 20 identical channels are used for the optical, blank
and temperature sensors.
 \label{fig3}}
\end{minipage}
\end{center}
\end{figure}

AC bridge circuits have been successfully used to read out pairs
of bolometers with DC stability to 30 mHz \cite{wilbanks}.  In
this scheme a pair of detectors is biased with an alternating
current so that resistance fluctuations are transformed into
changes in the AC bias amplitude across each detector.  These
signals are differenced in a bridge, amplified and demodulated.
The AC signal modulation eliminates the effects of 1/f noise in
the preamplifiers since the resulting signal spectrum is centered
about the carrier frequency.  The effects of drifts in the bias
amplitude, amplifier gain, and heat sink temperature are greatly
reduced if the two detectors in the bridge are well matched. The
optical responsivity of each the detectors in the bridge is
equivalent to that of a detector biased with the same rms DC power
and is constant in time as long as the average power on the
detector is constant over the course of one detector thermal time
constant.

For BOOMERanG the environmental background and the refrigerator
temperature $T_b$ are very stable, so we can employ an AC
stabilized "total power" readout system for individual bolometers
\cite{Hri}. This system contains a cooled FET input stage and
contributes less than 10 nV$_{rms}$/$\sqrt{\rm Hz}$ noise at all
frequencies within the bolometer signal bandwidth down to 20mHz.
The warm readout circuit has a gain stability of $< 10$
ppm/$^\circ$C , and we remove the large offset due to the
background power on the detector with a final stage high pass
filter with a cutoff frequency of 16 mHz. A block diagram of the
readout system is shown in fig.3.

There are many sources of RFI on the balloon that could couple to
bolometers. Microwave transmitters that send the data stream to
the ground and high current wires that drive the motors of the
Attitude Control System (ACS) are situated within a few meters of
the cryostat. The BOOMERANG wiring and focal plane are designed to
prevent RF interference from contributing to the noise of the
bolometers.

The bolometers are contained inside a 2~K Faraday cage inside the
cryostat.  RFI can enter the cryostat through the optical entrance
window and propagate into the 2~K optics box.  However, the exit
aperture of the optics box is RF sealed by the horn positioning
plate.  This plate contains feed horns with small waveguide
apertures for radiation from the sky to pass through to the
detectors. The largest waveguide feedthrough in this plate is 0.2"
in diameter which corresponds to a waveguide cutoff of 40~GHz.
Lower frequency RFI is reflected by this surface. Readout wires
entering the bolometer Faraday cage can also propagate RF signals
as coaxial cables. We run all of the bolometer wires through cast
ecosorb filters mounted to the wall of the Faraday cage to
attenuate these signals.  The filters are 30~cm long and have a
measured attenuation of <-20 dB at frequencies from 20~MHz to a
few~GHz.

The readout electronics are also sensitive to RFI.  We enclose all
of the cryostat electronics in an RF tight box that forms an
extension of the outer shell of the cryostat.  The signals from
the detectors pass through flexible KF-40 hose that is RF sealed
to the hermetic connector flange on the cryostat and to the wall
of the electronics box. The amplified signals exit the electronics
box through MuRata Erie RF filters mounted on the wall of the box.
We tested the RF sensitivity of the detectors with a RF sweep
generator covering the range 0.1-3~GHz.

\subsection{\it Optics}

The sub-degree angular resolution is obtained coupling the
bolometers feeds to an off-axis millimeter waves telescope
enclosed in a low emissivity cavity. A sketch of the optics is
shown in fig.4.

\begin{figure}[tb]
\begin{center}
\begin{minipage}[t]{16 cm}
\epsfig{file=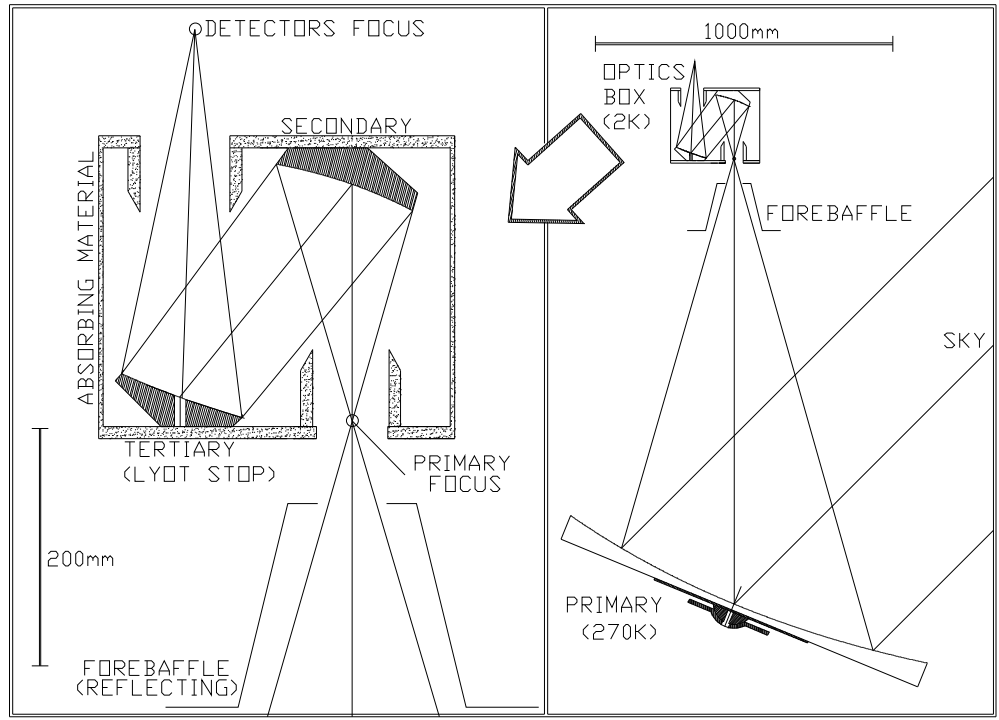,scale=1}
\end{minipage}
\begin{minipage}[t]{16.5 cm}
\caption{Optical system for BOOMERanG (see text).
 \label{fig4}}
\end{minipage}
\end{center}
\end{figure}

The BOOMERanG telescope consists of an ambient temperature 1.2~m
off-axis paraboloidal primary mirror which feeds a pair of cold
reimaging mirrors inside a large LHe cryostat.  The primary mirror
has a $45^{\circ}$ off- axis angle and can be tipped in elevation
by $+20^{\circ}$ and $-12^{\circ}$ to cover elevation angles from
$33^{\circ}$ to $65^{\circ}$. Radiation from the sky is reflected
by the primary mirror and passes into the cryostat through a thin
(0.002'') polypropylene window near the prime focus.  The window
is divided in two circles side by side, each 2.6" in diameter.
This geometry provides a wide field of view while allowing the use
of thin window material which minimizes the emission from this
ambient temperature surface.  Filters rejecting high frequency
radiation are mounted on the 77~K and 2~K shields in front of the
cold reimaging optics.  Fast off-axis secondary and tertiary
mirrors surrounded by black baffles reimage the prime focus onto a
detector focal plane with diffraction limited performance at 1~mm
over a $2^{\circ} \times 5^{\circ}$ field of view.  The reimaging
optics are also configured to form an image of the primary mirror
at the 10~cm diameter tertiary mirror.  The size of the tertiary
mirror therefore limits the illumination pattern on the primary
mirror which is underfilled by 50\% in area.

The BOOMERANG cold optics box shown in Fig.4 contains two
paraboloidal mirrors with effective focal lengths of 20~cm and
33~cm for the secondary and tertiary mirrors respectively.  The
parameters of these mirrors have been optimized for the required
performance with CodeV software. The input f/\# from the primary
mirror is f/2, so the focal plane is fed at f/3.3.  The tertiary
mirror is 10~cm in diameter, corresponding to an 85~cm diameter
aperture on the 1.2~m diameter primary mirror. All of the beams
overlap on the primary mirror by $>85$\%.  The large size of all
of the apertures in the system minimizes the effects of
diffraction.

The BOOMERANG optics are optimized for an array with widely
separated pixels.  The advantage of having large spacing between
pixels in the focal plane is the ability to difference the signals
from two such pixels and remove correlated optical fluctuations
such as temperature drift of the telescope, while retaining high
sensitivity to structure on the sky at angles up to the pixel
separation.  This scheme eliminates the need for moving optical
components and simplifies the design and operation of the
experiment.

The BOOMERANG focal plane contains a combination of single
frequency channels fed by conical horns and multicolor photometers
fed by winston horns.  Although the image quality from the optics
is diffraction limited over a $2^{\circ} \times 5^{\circ}$ field,
all of the feed optics are placed inside two circles $2^{\circ}$
in diameter, separated center to center by $3.5^{\circ}$.  The
focal plane area outside these circles is vignetted by blocking
filters at the entrance to the optics box and on the 77~K shield
and is unusable.  Due to the curvature of the focal plane, the
horns are placed at the positions of the beam centroids determined
from geometric ray tracing.  All of the feeds are oriented towards
the center of the tertiary mirror.

The projection on the sky of the BOOMERanG focal plane is shown in
fig.5.

\begin{figure}[tb]
\begin{center}
\begin{minipage}[t]{16 cm}
\epsfig{file=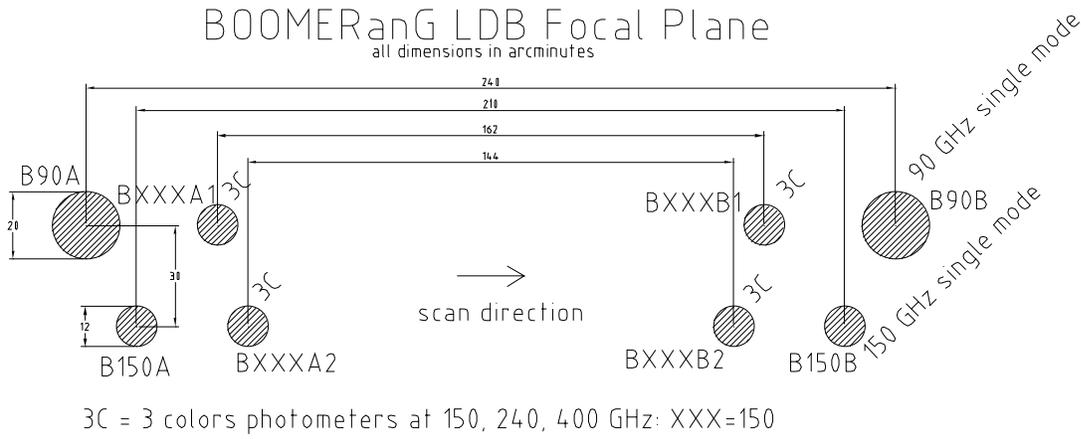,scale=0.8}
\end{minipage}
\begin{minipage}[t]{16.5 cm}
\caption{Projection on the sky of the BOOMERanG focal plane. All
the measurements are in arcminutes. The two larger beams are for
the single-mode 90GHz detectors, while the two detectors
immediately below are single-mode 150GHz detectors. The beams
A1..B2 are for the four multiband photometers, each observing
simultaneously at 150, 240 and 410 GHz. Slow azimuth scans ($\pm
30^o$, 1$^o/s$ or 2$^o/s$) are continuously performed at constant
elevation, while the center of the scan tracks the azimuth of the
lowest foreground region. A structure detected in the forward scan
in A1 will be detected a few seconds later in B1, and 30 seconds
later in time reverse in B1 and A1 during the return scan. Due to
sky rotation, the same sequence of events will be detected a few
minutes later in A2 and B2. Hours later, the same sky pixel is
observed again with a different inclination of the scan path. All
this is repeated every day for 10 days.
 \label{fig5}}
\end{minipage}
\end{center}
\end{figure}

The focal plane has been mapped from ground using a far-field
spherical black-body, and confirmed in flight through observations
of the compact HII region RCW38. The FWHM of the optical system
beams is $(18 \pm 2)'$ at 90GHz,  $(10 \pm 1)'$ at 150GHz, $(14
\pm 1)'$ at 240 GHz, and $(13 \pm 1)'$ at 410GHz. The effective
beam during the observations is obtained convolving the optical
beam with the pointing jitter, which is of the order of 2.5' rms
in the current pointing solution.

\subsection{\it Cryogenics}

The bolometers and the reimaging optics are cooled by a
combination of a large L$^4$He/LN cryostat and a high capacity
$^3$He fridge. The long hold time (2 weeks) has been obtained
using a 20K vapor cooled shield for the liquid He tank and
superinsulation for the nitrogen tank. The tanks are supported by
kevlar cords. The details of the main cryostat are reported in
\cite{Mas99}. The details of the $^3$He fridge are reported in
\cite{Mas98}. The cryogenic system performed well, keeping the
detectors well below their required operating temperature of 0.3~K
for the entire flight. A small daily oscillation of the main
helium bath temperature was induced by daily fluctuations of the
external pressure. In fact, the altitude of the payload varies
with the elevation of the sun, which oscillates between
11$^{\circ}$ to 35$^{\circ}$ diurnally.
 We have searched for scan
synchronous temperature fluctuations in the $^3$He evaporator and
in the $^4$He temperature, and we find upper limits of the order
of 1 $\mu K_{rms}$ during the 1$^{\circ}$/s scans, and drifts with
an amplitude of a few $\mu K$ during the 2$^{\circ}$/s scans.

\subsection{\it The payload}

A sketch of the payload is reported in fig.6. The total weight of
the science payload is $\sim$ 1.5 tons. There is one inner frame
supporting the telescope and the cryostat with the detection
system. The inner frame can tilt with respect to the outer frame,
in order to point the telescope at different elevations. The outer
frame supports the attitude control system (ACS), data storage and
telemetry electronics.

The ACS must be able to point a selected sky direction, and track
it or scan over it with a reasonable speed. The specifications are
1 arcmin rms for pointing stability, with a reconstruction
capability better than 0.5 arcmin. The BOOMERANG ACS is based on a
pivot which decouples the payload from the flight chain and
controls the azimuth, plus one linear actuator controlling the
elevation of the inner frame of the payload. The pivot has two
flywheels, moved by powerful torque motors with tachometers. On
the inner frame, which is steerable with respect to the gondola
frame, are mounted both the telescope and the cryogenic receiver.
The observable elevation range is between 33 and 65 deg. The
attitude sensors are a low and a high resolution sun sensor and a
set of laser gyroscopes (3-axis). A differential GPS is also used
as an absolute attitude reconstruction system. A flight programmer
CPU takes care of commands handling and observations sequencing; a
feedback loop controller CPU is used for digitization of sensors
data and PWM control of the current of the three torque motors.

\begin{figure}[tb]
\begin{center}
\begin{minipage}[t]{16 cm}
\epsfig{file=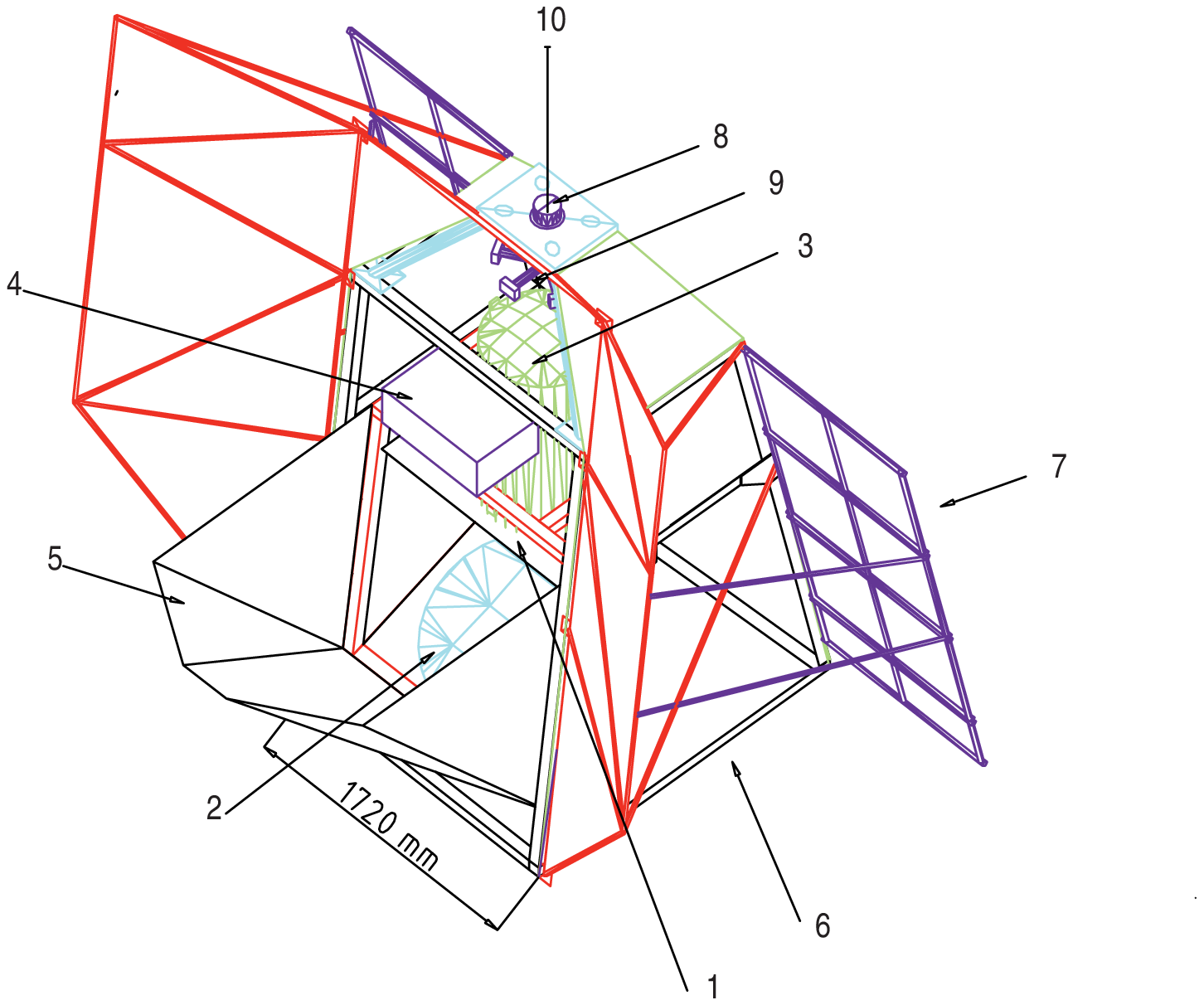,scale=0.8}
\end{minipage}
\begin{minipage}[t]{16.5 cm}
\caption{The BOOMERanG payload. On the inner frame (1) are mounted
the off axis primary mirror (2), the cryostat with the detection
system (3), the analog electronics (4), the ground shield (5). On
the outer frame (6) are mounted the sun shields, the solar panels
(7), the azimuth pivot (8) with the flywheel (9). The pivot
decouples the payload from the flight chain (10).
 \label{fig6}}
\end{minipage}
\end{center}
\end{figure}

\section{The LDB flight and the maps of the microwave-sky}

BOOMERanG was launched from McMurdo Station (Antarctica) on 29
December 1998, at 3:30 GMT. Observations began 3 hours later, and
continued uninterrupted during the 259-hour flight. The payload
approximately followed the 79$^\circ$ S parallel at an altitude
that varied daily between 37 and 38.5~km, returning within 50~km
of the launch site. The time-ordered data (TOD) comprises $5.4
\times 10^7$ 16-bit samples for each channel. These data are
flagged for cosmic-ray events, elevation changes, focal-plane
temperature instabilities, and electromagnetic interference
events. In general, about 5\% of the data for each channel are
flagged and not used in the subsequent analysis. The gaps
resulting from this editing are filled with a constrained
realization of noise in order to minimize their effect in the
subsequent filtering of the data. The data are deconvolved by the
bolometer and electronics transfer functions to recover uniform
gain at all frequencies; a high pass filter with null response at
$f=0$ and unit response at $f=0.01 Hz$ has been applied to remove
low frequency drifts and system instabilities. The pointing of the
telescope has been reconstructed for each data in the TOD
combining the signals of the gyroscopes, of the sun sensors and of
the differential GPS. A time ordered pointing (TOP) dataset has
been created for each pixel in the focal plane.

The gain calibrations are obtained from observations of the CMB
dipole. To compare with the data, we artificially sample the CMB
dipole signal \cite{Line1996}, corrected for the Earth's velocity
around the sun \cite{Stump1980}, according to the BOOMERANG
scanning, and filter this fake time stream in the same way as the
data. The 1~dps data is then fit simultaneously to this filtered
dipole, a similarly filtered dust emission model \cite{Fink99}, an
offset and the BOOMERANG 410 GHz data for all data more than 20
degrees below the Galactic plane. The dipole calibration numbers
obtained with this fit are robust to changes in Galactic cut, and
to whether or not a dust model is included in the fit; this
indicates that dust is not a serious problem for the
contamination. They are insensitive to the inclusion of a 410 GHz
channel in the fit, which is a general indication that there is no
problem with a wide range of systematics such as atmospheric
contamination, as these would be traced by the 410 GHz data. The
gain calibration accuracy is estimated to be $\sim 10 \%$, with
the error dominated by systematics.

\begin{figure}[tb]
\begin{center}
\begin{minipage}[t]{16 cm}
\epsfig{file=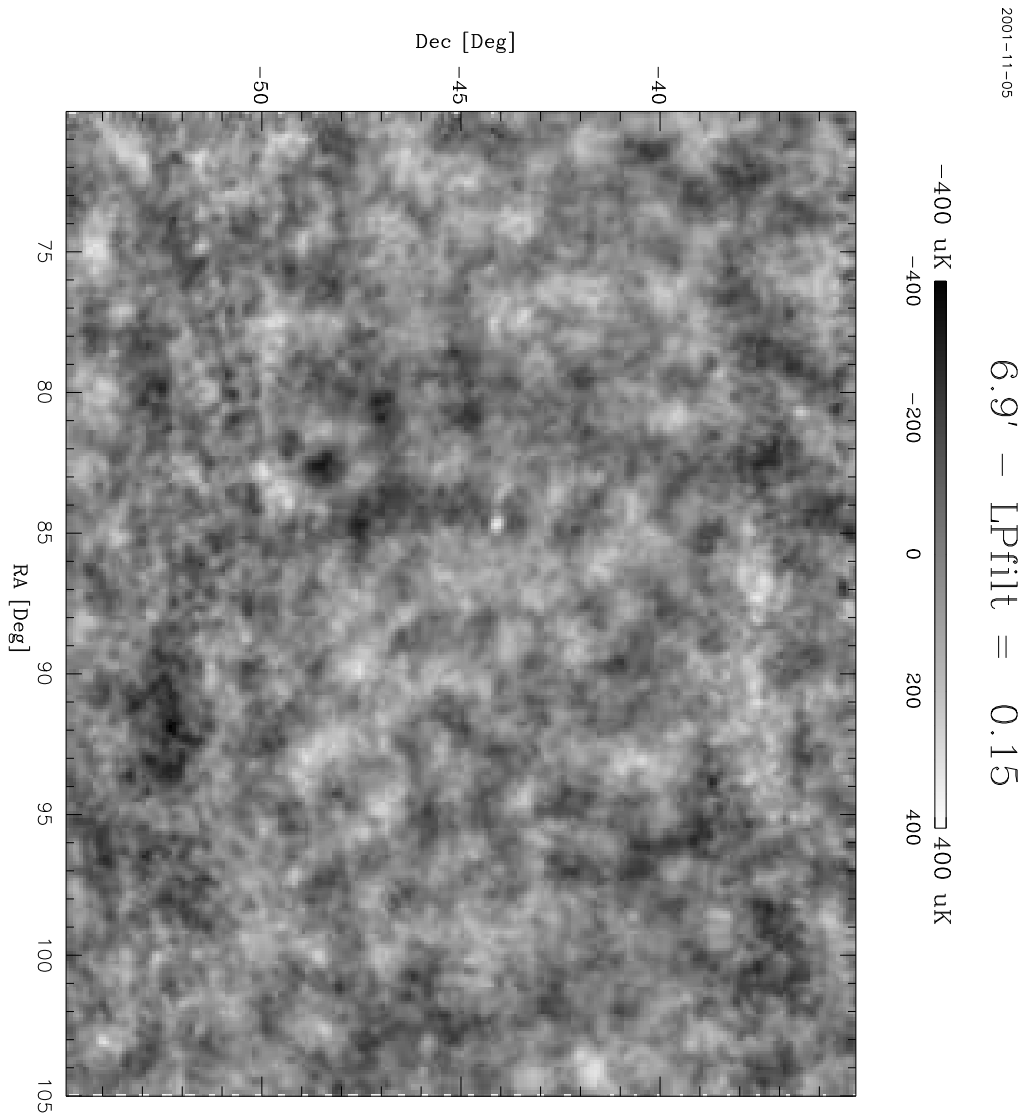,scale=0.7}
\end{minipage}
\begin{minipage}[t]{16.5 cm}
\caption{Map of the microwave sky measured by BOOMERanG at 150
GHz. The effective resolution of the observations is 12 arcmin.
 Healpix pixelization with 3.5' pixel side has been used.
 The map has been convolved with a 9' FWHM gaussian kernel for
 visualization purposes. The units correspond to
thermodynamic temperature fluctuations of a 2.73K blackbody. The
same structures are visible in the 90 GHz and in the 240 GHz maps
of BOOMERanG.
 \label{fig7}}
\end{minipage}
\end{center}
\end{figure}

Several independent methods have been used to estimate the map of
the observed sky region from the TOD and TOP: Naive maps (just
coadding data on the same pixel); maximum likelihood maps obtained
using the MADCAP package (\cite{Borr1999}); maximum likelihood
maps obtained using the iterative method of \cite{Nato2001};
suboptimal maps obtained using the fast map making method of
\cite{Hiv2001}. All methods produce very similar maps. In fig.7 we
show the central region of the 150GHz map from the B150A channel
obtained with the iterative method \cite{Nato2001}. The color code
used in the BOOMERanG maps corresponds to temperature fluctuations
of a 2.73K blackbody. Degree-scale structures with amplitude of
the order of 100 $\mu K$ are evident in the map at 150 GHz.
Consistent structure is also evident in the maps at 90 and 240
GHz. The similarity of the temperature maps obtained at different
frequencies \cite{debe2000} is the best evidence for the CMB
origin of the detected fluctuations.

\begin{figure}[tb]
\begin{center}
\begin{minipage}[t]{16 cm}
\epsfig{file=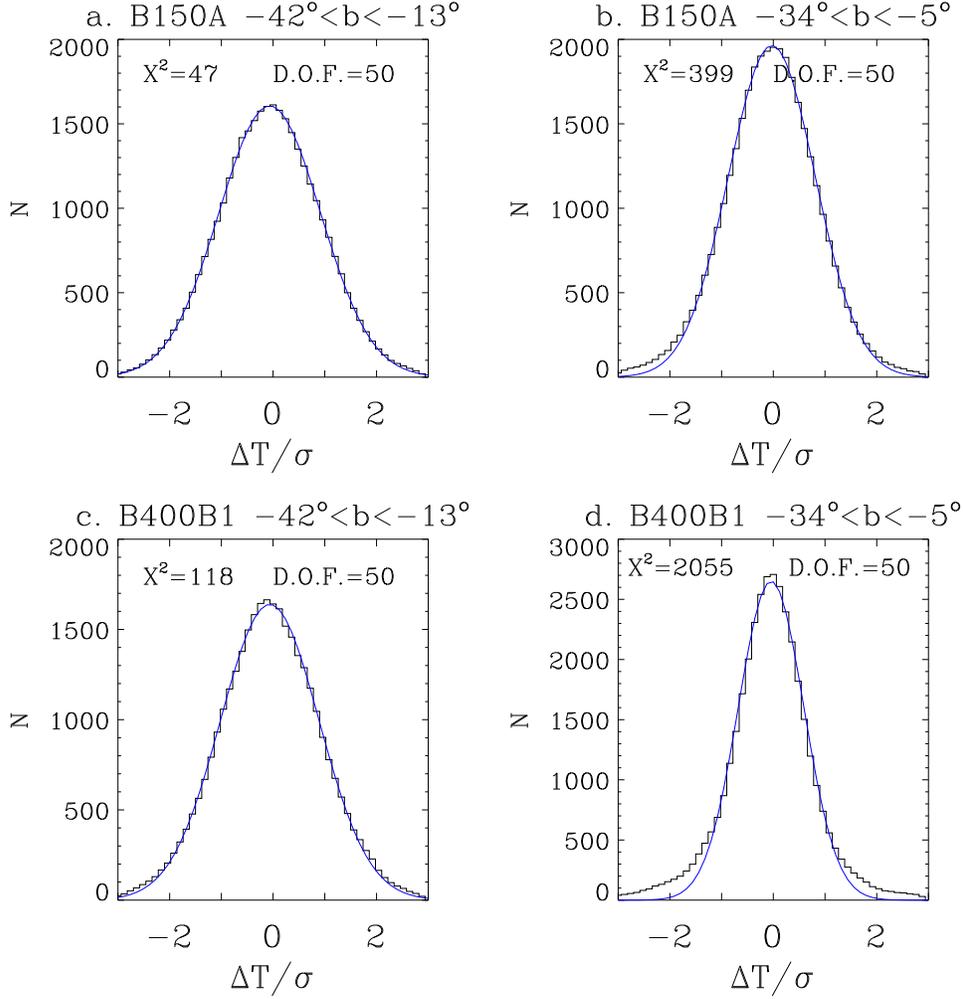,scale=0.7}
\end{minipage}
\begin{minipage}[t]{16.5 cm}
\caption{In the top row, we plot the 1-P distribution of the ratio
       $\Delta T_i / \sqrt{\sigma_i^2 + \sigma_{sky}^2}$ for the 150 GHz map
       (dominated by CMB fluctuations) in a box at high galactic latitude
       (a) and in a box at intermediate galactic
       latitudes (b). In the lower row the same distributions are plotted for
       the map at 410 GHz, which is dominated by interstellar dust
 \label{fig8}}
\end{minipage}
\end{center}
\end{figure}

Foregrounds contamination can be constrained significantly in the
center of the observed sky region \cite{debe2000},
\cite{Masi2001}. In fact, in the frequency region of interest, the
most important foregrounds are thermal emission from interstellar
dust and unresolved extragalactic sources. Galactic synchrotron
and free-free emission is negligible at this
frequency\cite{Kog99}. Contamination from extra-galactic point
sources is also small \cite{Tof98}; extrapolation of fluxes from
the PMN survey \cite{womb} limits the contribution by point
sources (including the three above-mentioned radio-bright sources)
to the angular power spectrum derived below to $<0.7\%$ at
$\ell=200$ and $<20\%$ at $\ell=600$. Masi et al. \cite{Masi2001}
have shown that thermal emission from interstellar dust, as mapped
by the IRAS satellite at 3000 GHz, is strongly correlated to the
emission dominating the BOOMERanG map at 410 GHz. The slope of a
linear fit of the BOOMERanG data vs the IRAS data is $S = (4700
\pm 1500) \mu K_{CMB}/(MJy/sr)$; Pearson's linear correlation
coefficient is R=0.138 with 68987 data (pixels) at Galactic
latitudes lower than -20$^o$. At this frequency, CMB anisotropy is
subdominant and barely detected in the BOOMERanG maps. On the
other hand, the lower frequency maps of BOOMERanG, where CMB
anisotropy is dominant, are much less correlated: the slopes are
$S = (258 \pm 52) \mu K_{CMB}/(MJy/sr)$ with R=0.041; $S=(46 \pm
29) \mu K_{CMB}/(MJy/sr)$ with R=0.003;  $S=(-20 \pm 110) \mu
K_{CMB}/(MJy/sr)$ with R=-0.028 at 240, 150 and 90 GHz
respectively. Using these data, it has been shown that the mean
square signal due to interstellar dust at 150 GHz is about two
orders of magnitude smaller than the CMB anisotropy. The residuals
of the correlation give upper limits of the same order of
magnitude for any other dust component (not correlated to the
emission mapped by IRAS).

\begin{figure}[tb]
\begin{center}
\begin{minipage}[t]{16 cm}
\epsfig{file=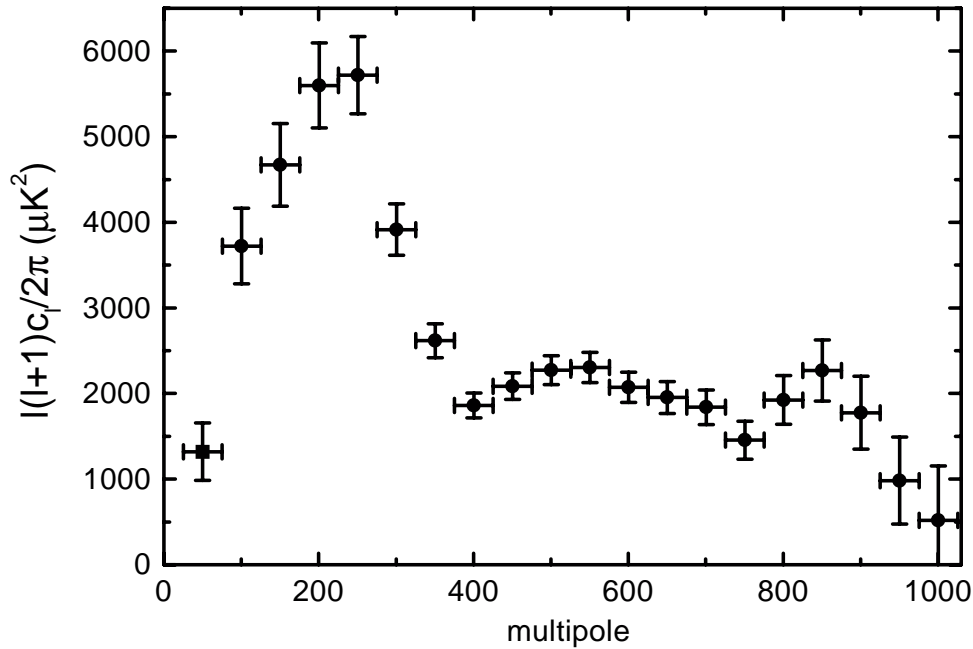,scale=0.8}
\end{minipage}
\begin{minipage}[t]{16.5 cm}
\caption{CMB anisotropy power spectrum detected by BOOMERanG at
150 GHz. The circles are from the Monte-Carlo approach, while the
point at $\ell = 50$ is from the rigorous maximum likelihood
approach. Only statistical errors (instrumental and cosmic/sample
variance) are shown. The gain calibration error is 10\% (i.e. 20\%
in the power spectrum), while the beam calibration error is 1.4'
in the FWHM, and affects (in a coherent way) the highest
multipoles.
 \label{fig9}}
\end{minipage}
\end{center}
\end{figure}

\section{Gaussianity}

According to the currently popular inflation scenario one
important property of the CMB anisotropy is its gaussianity. This
makes the CMB very special, since any other astrophysical source
produces strongly non-gaussian patterns in the sky. If the CMB is
really gaussian, all its statistical properties are encoded in its
angular power spectrum. At large scales, gaussianity of the CMB
has been investigated and confirmed using the COBE maps
\cite{Fer98, Band99, Kom01, Brom99}. At smaller scales, where the
natural tendency to gaussianity due to the central limit theorem
is less effective, the CMB has been found to be fully gaussian by
the QMASK \cite{Park01} and MAXIMA \cite{Sant01, Wu01}
experiments. Due to its wide sky and frequency coverage, BOOMERanG
is ideally suited to carry out an accurate analysis of the
possible systematic effects present in the detected signal. This
issue is studied in detail in \cite{Pole02}, who use five
estimators of departures from gaussianity: the skewness and
kurtosis of the CMB temperature distribution, and the three
Minkowski functionals of the maps: area, length and genus. All
these methods confirm the gaussianity of the CMB fluctuations
detected by BOOMERanG. In fig.8 we plot the simple 1-point
distribution of the data: the gaussianity of the CMB dominated
high latitude 150 GHz map is evident, while the sky is not
gaussian at lower latitudes and at 410 GHz (where emission is
contaminated or dominated by interstellar dust and detector
noise).

\section{The Power spectrum}

We have computed the power spectrum of the 150 GHz maps detected
by BOOMERanG with two independent methods: the rigorous (and
computationally expensive) maximum likelihood approach of
\cite{Borr1999}, and the Monte-Carlo approach of \cite{Hivo2001}.
Both methods give consistent results. The best estimate of the
power spectrum from BOOMERanG at 150 GHz is shown in fig.9, where
we have combined the results of the Monte Carlo method of
\cite{Nett2001} at $\ell > 75$ with the lowest bin of the maximum
likelihood spectrum of \cite{debe2000}, corrected for the improved
calibration. In fig.10 we compare the BOOMERanG power spectrum at
150 GHz to the spectra measured by the experiments DASI (at 30
GHz)\cite{dasi} and MAXIMA (at 150 GHz) \cite{lee}. Despite of the
orthogonal experimental techniques and different observed regions,
the three results are statistically consistent.

Netterfield et al. \cite{Nett2001} have discussed how the measured
power spectrum of the sky is robust against variations of the
$\ell$-binning, channel selection, data subset selection, effects
of uncertainties in the beam and effects of the noise. de
Bernardis et al. \cite{debe2002} have shown that the three peaks
and two dips present in the power spectrum are statistically
significant. The first peak is at $\ell_1 = (213^{+10}_{-13})$
(the errors correspond to a 1$\sigma$ confidence interval in the
location if the peak).  Its amplitude is detected at $\simgt 5
\sigma$, while for the second peak (at $\ell_2 =
(541^{+20}_{-32})$) and third peak (at $\ell_3 =
(845^{+12}_{-25})$) the amplitudes are detected at basically
2$\sigma$. Several methods to measure the location and amplitude
of the peaks have been compared, all producing very consistent
results. In particular, the results of fits using empirical
functions are consistent with the results of fits using a database
of adiabatic inflationary spectra of the CMB \cite{debe2002}.

\begin{figure}[tb]
\begin{center}
\begin{minipage}[t]{16 cm}
\epsfig{file=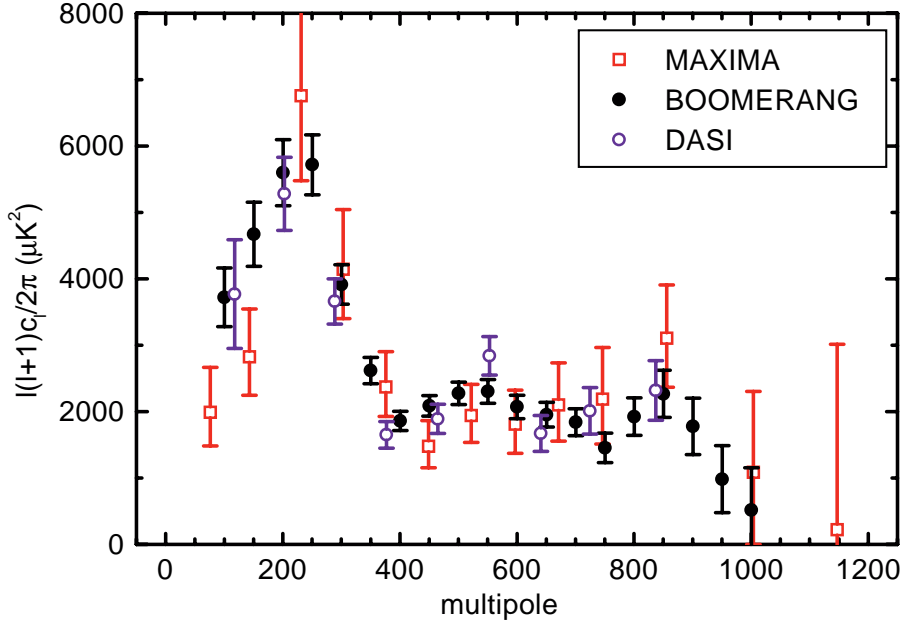,scale=0.8}
\end{minipage}
\begin{minipage}[t]{16.5 cm}
\caption{CMB anisotropy power spectrum detected by BOOMERanG,
MAXIMA and DASI. Approximately uncorrelated band-powers are
plotted for each of the experiments. The error bars represent
statistical errors only.
 \label{fig10}}
\end{minipage}
\end{center}
\end{figure}

\section{Cosmological implications}

The measurement of $\ell_1$ from the most recent BOOMERanG power
spectrum is $\ell_1 = (213^{+10}_{-13})$, which is consistent with
a flat geometry of the Universe. Rigorous confidence intervals for
the parameter $\Omega$ can be found with the Bayesian analysis of
the full power spectrum, as described below. These methods use the
full information content of the data (and not just the location of
the peak) and take in due account the degeneracies between
different parameters.

The ratio between the amplitude of the second peak and the
amplitude of the first one depends mainly on the physical density
of baryons $\Omega_b h^2$ and on the tilt of the density
fluctuations spectrum. A high density of baryons favors
compressions against rarefactions: the odd-order peaks
(compression) are enhanced while the even-order peaks
(rarefaction) are depleted. From the BOOMERanG power spectrum the
ratio is $(5450 \pm 350) / (2220 \pm 330) = (2.45 \pm 0.52)$.
Assuming a scale invariant power spectrum of the density
fluctuations ($n=1$), this corresponds to a physical density
$\Omega_b h^2 \sim 0.02$. Again, better constraints are found by
means of the Bayesian analysis of the full power spectrum as
described below. In fact, a tilt of the density fluctuations
spectrum ($n < 1$) has the same effect of a high baryons density
in depleting the second peak with respect to the first one, so
there is a degeneracy between the two parameters. But the effects
of the two quantities on the amplitude of the third peak are
different. The amplitude of the third peak is increased by a high
baryon density, while it is decreased by a red ($n<1$) primordial
density fluctuation spectrum. The result is that extending the
observations to $\ell \sim 1000$ breaks the degeneracy between $n$
and $\Omega_b h^2$, thus allowing a determination of both the
parameters.

\begin{figure}[tb]
\begin{center}
\begin{minipage}[t]{16 cm}
\epsfig{file=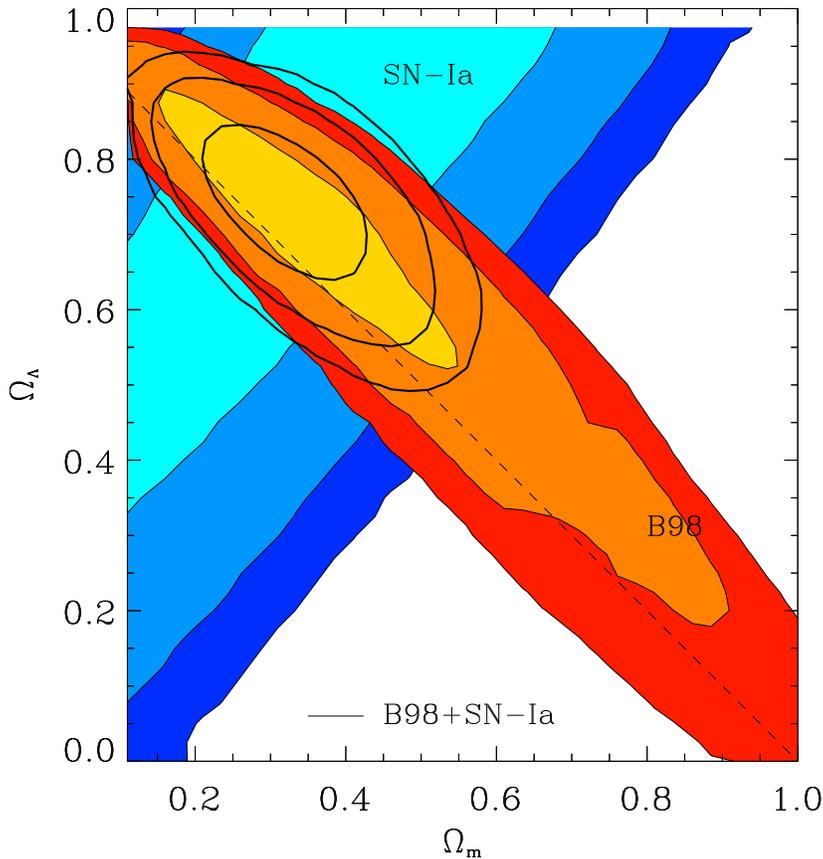,scale=0.7}
\end{minipage}
\begin{minipage}[t]{16.5 cm}
\caption{Constraints in the $\Omega_m$ vs. $\Omega_\Lambda$ plane
from the combined BOOMERANG and COBE/DMR datasets, assuming the
weak prior  $0.45 < h < 0.85$, and an age of the Universe $> 10$
Gyr. The likelihood at each point is calculated by maximizing over
the remaining cosmological parameters. The shaded regions
correspond to the $68.3\%$, $95.4\%$ and $99.7\%$ confidence
levels. The CMB contours (labeled B98) are overlaid on the
contours derived from observations of high redshift supernovae
(labeled SN 1a). The line contours are for the combined
likelihood.
 \label{fig11}}
\end{minipage}
\end{center}
\end{figure}

It is important to stress the fact that our result for $\Omega_b
h^2$ agrees with the constraint on $\Omega_b h^2$ from the Big
Bang Nucleosynthesis. In fact, the physical density of baryons
affects the yield of the nuclear reactions happening in the first
few minutes after the big bang. The resulting primordial
abundances of light elements are measured by the optical
absorption spectra of primordial clouds of matter \cite{burl2000}.
It is evident that both the physics and the experimental methods
involved in these two measurements of $\Omega_b h^2$ are
completely orthogonal to the CMB ones. The fact that the two
estimates of $\Omega_b h^2$ agree so well should be considered a
great success of the Hot Big Bang model.

The multiple peaks and dips are a strong prediction of the
simplest adiabatic inflationary models, and more generally of
models with passive, coherent perturbations. Although the main
effect giving rise to them is regular sound compression and
rarefaction of the photon-baryon plasma at photon decoupling,
there are a number of influences that make the regularity only
roughly true. The best way to extract all the information encoded
in the data is by comparison to a large database of $C_\ell$
spectra. In order to limit the size of the database, we considered
for the first approach the class of adiabatic inflationary models.
We have explored a parameter space with 6 discrete parameters and
a continuous one. The parameters ranged as follows: $\Omega_{m}=
0.11, ..., 1.085$, in steps of 0.025; $\Omega_{b} = 0.015,
...,0.20$, in steps of 0.015; $\Omega_{\Lambda}=0.0, ..., 0.975$,
in steps of 0.025; $h=0.25, ..., 0.95$, in steps of 0.05; spectral
index of the primordial density perturbations $n_s=0.50, ...,
1.50$, in steps of 0.02, $\tau_C = 0., .., 0.5$, in steps of 0.1.
The overall amplitude $C_{10}$, expressed in units of
$C_{10}^{COBE}$, is allowed to vary continuously.  We used the
BOOMERANG power spectrum expressed as $18$ bandpowers $C_b$
\cite{Nett2001} and we computed the likelihood for the
cosmological model $C_b^T$ as $\exp(-\chi^2 /2)$, where $\chi^2 =
(C_b - C_b^T) M^{-1}_{bb^\prime} (C_{b^\prime} - C_{b^\prime}^T)$.
Here $M_{bb^\prime}$ is the covariance matrix of the measured
bandpowers; $C_b^T$ is an appropriate band average of $C_\ell$. A
$10 \%$ Gaussian-distributed calibration error in the gain and a
1.4' (13\%) beam uncertainty were included in the analysis as
additional parameters with gaussian priors. The COBE-DMR
bandpowers used were those of \cite{BJK98}, obtained from the
RADPACK distribution \cite{Radpack}. The 95\% confidence intervals
for the parameters we find in this way depend to some extent on
the priors assumed. Using COBE and BOOMERanG data only, with a
weak prior $0.45 < h < 0.9$ significantly constrains three
parameters: $0.9 < \Omega < 1.15$, $0.8 < n < 1.1$  and $0.015 <
\Omega_b h^2 < 0.029$.

The detailed breakdown of $\Omega$ in $\Omega_m$ and
$\Omega_\Lambda$ cannot be inferred by CMB data alone. In fig.11
we plot the joint likelihood for $\Omega_m$ and $\Omega_\Lambda$
using the BOOMERanG and COBE data, and the weak prior $0.45 < h <
0.85$ and an age of the Universe $>$ 10 Gyrs.

Using more restrictive priors, deriving from the properties of the
large scale distribution of Galaxies ($\sigma_8$ and $\Gamma$), or
the data of high-redshift supernovae, or the measurement of $h$ by
the HST, produces narrower, consistent intervals for all the
parameters \cite{Nett2001}, \cite{debe2002}. This fact suggests a
good overall consistency of the present cosmological paradigm.
Including these priors, it is also possible to constrain the two
additional forms of mass-energy contributing to the total
mass-energy density in the Universe, i.e. dark matter and dark
energy. We find that the 95\% confidence intervals for $\Omega_m
h^2$ and $\Omega_\Lambda$ are $0.36 < \Omega_\Lambda < 0.72$ and
$0.09 < \Omega_m h^2 < 0.18$ (LSS prior); $0.52 < \Omega_\Lambda <
0.88$ and $0.01 < \Omega_m h^2 < 0.17$ (SN1a prior); $0.40 <
\Omega_\Lambda < 0.84$ and $0.06 < \Omega_m h^2 < 0.26$ (HST $h$
prior). The detection of a non-zero $\Omega_\Lambda$ comes thus
from independent paths and sets a formidable challenge to our
understanding of fundamental physics \cite{Wei89}.

\section{Conclusions}

The BOOMERanG experiment has produced multi-frequency maps of the
microwave sky, where the structure of the CMB has been resolved
with high signal to noise ratio. The structures in the CMB are
gaussian, and their power spectrum features three peaks. This is
consistent with the presence of acoustic oscillations in the
primeval plasma. It also fits the predictions of the adiabatic
inflationary scenario. The values of the cosmological parameters
inferred in this scenario point to a flat universe (the 95\%
confidence interval is $0.85 < \Omega < 1.1$) with nearly
scale-invariant initial adiabatic perturbations and a significant
contribution of dark energy to the total density of the Universe.
These results from BOOMERanG have been confirmed by independent
CMB experiments (like DASI and MAXIMA) and by other cosmological
observations. The forthcoming space missions MAP and Planck will
improve significantly the precision of these results, either
entering the "precision cosmology" era, or detecting hidden
inconsistencies of the present cosmological scenario.

\end{document}